\documentclass[twoside]{article}
\usepackage[accepted]{aistats2e}
\usepackage{graphicx}
\usepackage{amssymb}
\usepackage{amsmath}
\usepackage{amsthm}
\usepackage{epstopdf}
\usepackage{multirow}
\usepackage{rotating}
\newcommand{\be}{\begin{eqnarray} \begin{aligned}}
\newcommand{\ee}{\end{aligned} \end{eqnarray} }
\newcommand{\benn}{\begin{eqnarray*} \begin{aligned}}
\newcommand{\eenn}{\end{aligned} \end{eqnarray*} }

\usepackage{color}
\usepackage{url}

\usepackage{bm}
\newcommand{\vc}[1]{{\bm #1}}	

\theoremstyle{plain}
\newtheorem{theorem}{Theorem}[section]

\numberwithin{equation}{section}
\newtheorem{defn}{Definition}[section]

\begin{document}

\twocolumn[

\aistatstitle{Statistical Tests for Contagion in Observational Social Network Studies}

\aistatsauthor{ Greg Ver Steeg \and Aram Galstyan }

\aistatsaddress{ USC Information Sciences Institute\\ Marina del Rey, CA 90292, USA \\ \{gregv,galstyan\}@isi.edu } ]



\begin{abstract}
Current tests for contagion in social network studies are vulnerable to the confounding effects of latent homophily (i.e., ties form preferentially between individuals with similar hidden traits). We demonstrate a general method to lower bound the strength of causal effects in observational social network studies, even in the presence of arbitrary, unobserved individual traits. Our tests require no parametric assumptions and each test is associated with an algebraic proof. We demonstrate the effectiveness of our approach by correctly deducing the causal effects for examples previously shown to expose defects in existing methodology. Finally, we discuss preliminary results on data taken from the Framingham Heart Study. 
\end{abstract}



Christakis and Fowler's paper suggesting that obesity may spread along social ties~\cite{obesity} has sparked years of discussion about what constitutes evidence of contagion in observational social network studies (see, e.g., this recent review~\cite{christakis2012}). 
The most general result from the causal modeling perspective shows that latent homophily acts as a confounder for contagion so that uniquely pinpointing the strength of contagion is impossible without additional assumptions~\cite{cosma}. In other words, contagion is non-parametrically unidentifiable. 
However, if the true goal is to test for the presence of contagion, a lower bound on the strength of contagion is all that is necessary. We present a general method to obtain such bounds in this paper. 
%

Identifying causal effects in social networks through intervention is often impractical or even unethical. Measuring all the human traits that affect link formation and observed actions is unrealistic. A method to measure the strength of causal effects without recourse to these alternatives is of central importance for studying social networks.
Our method produces a sequence of bounds on the strength of contagion which converge in some limit to the best possible bounds. In this sense, our method is the best solution to the problem of measuring the strength of contagion that does not involve invoking additional (parametric) assumptions.


%
%
%
%
\section{Model and Method}\label{sec:theory}

\begin{figure}[pt]
\vspace{0.6cm}
\center
\includegraphics[width=4.5cm]{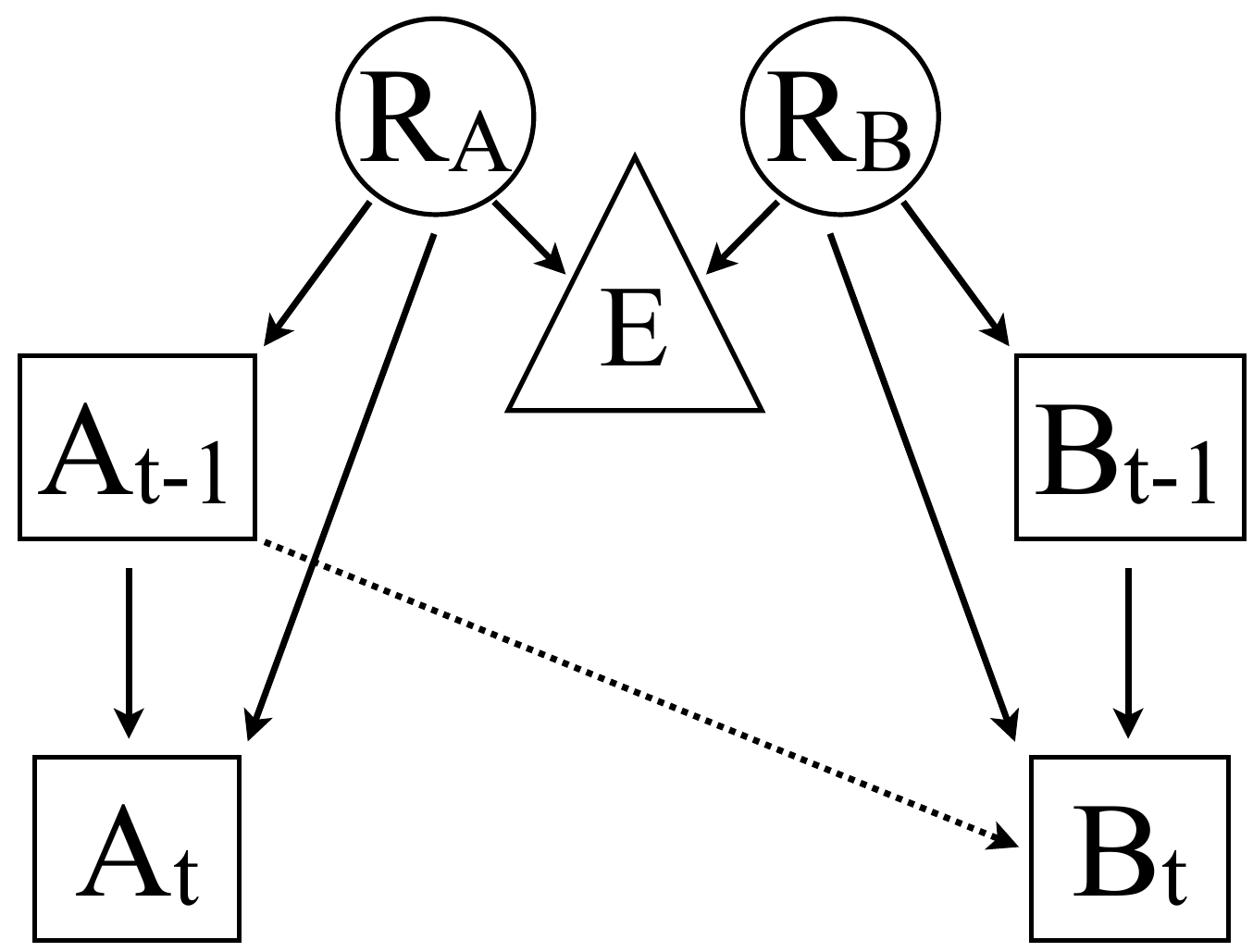}
\caption{A slice of a Bayesian network representing both latent homophily and contagion (dotted line). We observe a sequence of actions for $A$ and $B$ that depend on previous actions and on hidden attributes $R_A,R_B$. This graph is conditioned on the presence of a directed edge in the social network, $E$, from $A$ to $B$, whose formation depends on $R_A,R_B$.}
\label{fig:homophily}
\end{figure}

We have two actors Alice(A) and Bob(B) whose actions or attributes we observe at discrete time steps, $t=1,\ldots,T$. 
In Fig.~\ref{fig:homophily}, we depict a Bayesian network that incorporates both contagion (also known as social influence) and homophily, following Shalizi and Thomas \cite{cosma}. The only difference is that we are more pessimistic in that we will consider all the attributes of Alice($R_A$) and Bob($R_B$) to be hidden. In this figure, we condition on $E$, the presence of a
directed edge from Alice to Bob. The formation of such an edge depends in some arbitrary way on the hidden attributes of Alice and Bob. We often refer to this process as {\em latent homophily}, even though the edge formation process is unrestricted and could, e.g., be heterophilous instead (i.e. edges are more likely to form between actors with different attributes). 

We observe some sequence of actions $(A_1,\ldots,A_T)$ (sometimes abbreviated $A_{1:T}$ or simply $A$) and $B_{1:T}$. 
Given $E$, what correlations are possible between $A$ and $B$? Below we use standard results about graphical models \cite{pearl} for the network in Fig.~\ref{fig:homophily} along with some simple manipulations using Bayes' rule.  We also employ the common shorthand that capital letters represent random variables and we suppress their instantiation when no ambiguity arises, i.e., $ P(A_1) \equiv P(A_1 = a_1) $. 
We additionally require that the transitions are stationary, i.e., $\forall t,t', ~~P(A_t | A_{t-1}, R_A)=P(A_{t'} | A_{t'-1}, R_A)$, and similarly for $B$. 
In principle, we could allow the actions for Alice and Bob to come from any finite discrete set, but for simplicity we will consider $A_t,B_t \in \{0,1\}$ from here on. Surprisingly, we can allow the size of the hidden attribute space to be infinite. The probability distribution over $A,B$, conditioned on the presence of an edge, $E$, takes the following form.
\begin{align}\label{eq:jointprob}
\lefteqn{P(A_{1:T},B_{1:T}|E) =}\\
&&  \sum_{R_A,R_B} P(R_A,R_B | E) P(A_1 | R_A) P(B_1| R_B) \times ~~ \nonumber\\
 &&  \prod_{t=2}^T  P(A_t | A_{t-1}, R_A) P(B_t | B_{t-1},A_{t-1}, R_B) \nonumber
\end{align}

The Bayesian network in Fig.~\ref{fig:homophily} represents a class of models that can be specified by setting the conditional probability distributions for each node, $P(X | \mbox{parents}(X))$. In this case, unidentifiability means that a particular (non-experimental) probability distribution over observed variables does not pick out a unique model. Therefore, the strength of the dependence of $B_t$ on $A_{t-1}$ is also not uniquely determined. In technical terms the presence of a back-door path from $A_{t-1}$ to $B_t$ is a confounder preventing a unique identification of the causal effect of $A_{t-1}$ on $B_t$ (see \cite{cosma,pearl}). Our goal is not a unique identification of the strength of the causal effect, but rather to establish a lower bound on the strength.
We begin by slicing up the space of all possible models according to the strength of the causal effect.

 \begin{defn}\label{def}
We define $\Delta$-causal models for $\Delta \equiv [\delta_l,\delta_u]$ as the set $\mathcal P_{\Delta}$, consisting of probability distributions, $P(A,B|E)$, s.t. there exists conditional probability distributions that satisfy Eq.~\ref{eq:jointprob} and that additionally satisfy:
\begin{align}\label{eq:bound}
\lefteqn{ \delta_l \leq P(B_t = 1 | B_{t-1},A_{t-1} = 1, R_B)} \\
 &&- P(B_t = 1 | B_{t-1},A_{t-1} = 0, R_B) \leq \delta_u, \nonumber
\end{align}
for all possible values of $B_{t-1}, R_B$.

The class of models specified by $\delta_l=\delta_u=0$ we denote by $\mathcal P_0$ and refer to as non-causal models.
\end{defn}
The quantity in Eq.~\ref{eq:bound} is conventionally referred to as the {\em average causal effect of treatment}, or just average treatment effect, where the treatment in this case refers to Alice's action and the effect is measured on Bob \cite{morganbook}. We are really bounding the average treatment effect for every sub-population defined by $R_B$. 
Identifying whether a distribution is in the set of non-causal models is of special interest and because $\delta_l=\delta_u=0$, this implies that $P(B_t | B_{t-1},A_{t-1}, R_B)$ simplifies to $P(B_t | B_{t-1}, R_B)$ in that case.  

\subsection{Simple Example}
Consider a simple function of the observed variables $c(A_{1:T},B_{1:T})$, or $c(A,B)$. The expectation value of this function is 
$$ \langle c(A,B) \rangle_P \equiv \sum_{A,B \in \{0,1\}^T} P(A,B|E) c(A,B).$$

Set $T=4$ and consider a specific observable, 
\begin{align}\label{eq:c1}
\lefteqn{c^{(1)} (A_{1:T},B_{1:T}) =} \\
& & \left(\mathbf{1}\{A_2 = B_2 \neq A_3 = B_3\}  -  \mathbf{1}\{A_2 = B_3 \neq A_3 = B_2\} \right) \nonumber\\
&& \times (1-  \mathbf{1}\{A_1 \neq A_4\} \mathbf{1}\{B_1 \neq B_4\}) \nonumber
\end{align}
This operator can only take values $0$ or $\pm 1$, so its average must lie in this range. Using Def.~\ref{def} and simple but tedious algebra verifies that
\be\label{eq:c1b}
\forall P \in \mathcal P_0,~~ \langle c^{(1)}(A,B) \rangle_P = 0.
\ee
While a fact like this is straightforward to verify, it offers little understanding. In the rest of the paper, we develop methods to find equalities (and inequalities) of this form. Moreover, we focus the search by looking for useful tests so that, e.g., we find conditions that are satisfied $\forall P \in \mathcal P_0$, but are violated by models which contain contagion. 

For instance, if we define a simple model of influence, $ P_\delta (A,B)$ in which $P(A_t=0) =P(A_t=1)=P(B_1=0) =P(B_1=1)= 1/2$ and $B_t = A_{t-1}$ with probability $\delta$ otherwise $B_t$ randomly becomes $0$ or $1$. The ``average treatment effect'' in this case is just $\delta$. We can easily see that 
$$ \langle c^{(1)}(A,B) \rangle_{\hat P_\delta} = -\frac{3}{16} \delta.$$
Even for a tiny amount of influence Eq.~\ref{eq:c1b} is violated, demonstrating that the distribution $P_{\delta} (A,B)$ cannot be explained by a non-causal model, even with an infinite number of hidden attributes. 

\subsection{Finding Useful Tests}
Determining if a (non-experimental) probability distribution is compatible with a class of models defined by Def.~\ref{def} seems hopeless because the number of parameters depends on the size of the hidden attribute space which can be infinite. Luckily, as we have just seen, we can find simple conditions which all distributions in $\mathcal P_\Delta$ satisfy. A distribution that violates one of these conditions is incompatible with the associated class of models. 

We begin by considering a candidate probability distribution, $\hat P(A,B)$, a class of models (specified by a set of distributions) $\mathcal P$, and some observable, $c(A,B)$. 
We are looking for the following condition to be satisfied.
\begin{align*}
\langle c(A,B) \rangle_{\hat P} - \langle c(A,B) \rangle_{P} \geq \gamma > 0, \quad \forall P \in \mathcal P
\end{align*}
If this condition is satisfied then $c(A,B)$ constitutes a statistical test that is bounded $\forall P \in \mathcal P$, but is violated for the distribution $\hat P$. Looking for an observable $c(A,B)$ with associated bound $\gamma$ leads us to an optimization problem.
\begin{gather}\label{eq:opt1}
 \underset{\gamma,c(A,B)}{\text{maximize }}  \gamma, \text{  s.t.} \\
 -\gamma +\langle c(A,B) \rangle_{\hat P} - \langle c(A,B) \rangle_{P} \geq 0, \quad \forall P \in \mathcal P \nonumber
\end{gather}
Our goal is to transform this optimization problem into a sequence of linear programs (LP), so that the lower bound, $\gamma$ becomes successively tighter as we increase the size of the LP. To that end, we will first represent the expectation values in terms of polynomials. This allows us to represent the condition in the second line using a result about representations of non-negative polynomials which we include here.

\begin{theorem}\label{handelman}
(Handelman's representation \cite{handelman})
Any polynomial, $\gamma - h(x)$, that is positive on a compact domain $\mathcal K = \{x: g_1 (x) \geq 0,\ldots, g_s(x) \geq 0\}$, where the $g_i(x)$ are linear, can be written in this form,
$$
\gamma - h(x) = \sum_{k \in \mathbb N^s} \lambda_k \prod_{i=1}^s g_i(x)^{k_i},
$$
using non-negative $\lambda$'s, with $k$ representing a vector of non-negative integers.
\end{theorem}
Because the RHS consists of a sum of products of non-negative quantities on $\mathcal K$, we can see the LHS should be non-negative. The theorem ensures any positive polynomial can be written in this form. 
The main drawback, however, is that it does not say how many terms are required. Although this shortcoming is remedied in \cite{convergence}, those bounds are often impractical. Instead, we can bound the number of terms so that $\sum_i k_i \leq d_{max}$, and therefore $\gamma$ is an upper bound for $h(x)$ on $\mathcal K$, that becomes progressively tighter as we increase $d_{max}$.
As a bonus, providing a concrete representation in terms of $\lambda$'s provides a certificate, or an algebraic proof, that $\gamma$ is an upper bound for $h(x)$ on $\mathcal K$ (see Sec.~\ref{sec:bounds} for an explicit example).

Now we can proceed to re-write the optimization problem in Eq.~\ref{eq:opt1} as an LP using Handelman's representation. First, looking at Eq.~\ref{eq:jointprob} and using convexity, we see that, for $P\in \mathcal P_\Delta$
\begin{align}\label{eq:convexity}
\lefteqn{\langle c(A,B) \rangle_{P \in \mathcal P_\Delta} \leq \max_{P' \in \mathcal P_\Delta} \langle c(A,B) \rangle_{P'} =} \\
&&\max_{P(\cdot | \cdot)} \sum_{A, B \in \{0,1\}^T} c(A,B) P(A_1 | R_A) P(B_1| R_B) \times  \nonumber\\
 &&~~\prod_{t=2}^T  P(A_t | A_{t-1}, R_A) P(B_t | A_{t-1},B_{t-1}, R_B) \nonumber
\end{align}
The maximization is over conditional probability distributions that satisfy normalization, positivity, and the condition in Eq.~\ref{eq:bound}. We think of the conditional probability distributions as variables, e.g. $x_1 \equiv P(A_1 = 0 |R_A)$, normalization is ensured by writing $P(A_1 = 1 |R_A) = 1 - x_1$, positivity corresponds to conditions like $g_1(x) = x_1 \geq 0, g_2(x) = 1-x_1 \geq 0,\ldots$, and Eq.~\ref{eq:bound} corresponds to more complicated linear inequalities involving these variables that depend on $\delta_l,\delta_u$. We represent all these linear inequalities with the set $\mathcal K_\Delta$. 
We will give a more concrete demonstration of this mapping in the next section. 

To complete the transformation of Eq.~\ref{eq:convexity}, we also consider the vector of variables, $\vc c$, whose elements we will sometimes index $c_{AB} \equiv c(A,B)$, where the concatenated binary sequences $A$ and $B$ should be interpreted as an integer in $[0,2^{2T}-1]$. We can do the same to represent $\hat P(A,B)$ as a vector $\hat{\vc p}$, and then expectation values are just dot products. 
Putting this together, we re-write this equation as
\begin{align*}
\langle c(A,B) \rangle_{P \in \mathcal P_\Delta} \leq \max_{\vc x \in \mathcal K_\Delta} \vc c \cdot \vc f(\vc x) \nonumber
\end{align*}
We can ensure $-\gamma + \langle c(A,B) \rangle_{\hat P}$ is an upper bound for the RHS (which is the condition written in the second line of Eq.~\ref{eq:opt1}) by ensuring that $-\gamma + \vc c \cdot \hat{\vc p} - \vc c \cdot \vc f(\vc x)$ has a Handelman representation (and is therefore non-negative). 

That leads us to the following form of the optimization in Eq.~\ref{eq:opt1}.
\begin{gather}\label{eq:LP1}
 \underset{\gamma,\vc c,\vc \lambda}{\text{maximize }}  \gamma, \text{  s.t.} \\
-\gamma +\vc c \cdot \hat{\vc p} - \vc c \cdot \vc f(\vc x) = \sum_{k \in \mathbb N^s} \lambda_k \prod_{i=1}^s g_i(x)^{k_i} \nonumber \\
\gamma, \vc \lambda \geq 0, c_i \in [-1,1], \sum_{i} k_i \leq d_{max}\nonumber
\end{gather}
Equating the terms of the polynomials on both sides of the second line results in linear equalities among the variables. We restrict the $c_i$'s to some fixed range so that $\gamma$ cannot be made arbitrarily large by scaling all the $c_i$'s. This optimization turns out to be a linear program (LP), and, hence can be efficiently solved in polynomial time. 
The feasibility of this LP proves that there is a linear equality that is obeyed by all distributions in $\mathcal P_\Delta$ but is violated by the distribution $\hat P$. Namely, we have shown that
$$\forall P \in \mathcal P_\Delta, \quad \langle c(A,B) \rangle_P \leq \langle c(A,B) \rangle_{\hat P} - \gamma.$$
So obviously if $\hat P \in \mathcal P_\Delta$ this would lead to a contradiction (assuming $\gamma$ is positive). 

Not only does this LP provide us with a concrete bound and the size of the violation by $\hat P$, the $\lambda$'s can be interpreted as an algebraic proof of the upper bound. The main factor determining the size of the LP is the number of variables, $\lambda_i$, which is determined by the number of terms we use in our Handelman representation.
Mathematica can solve LPs with hundreds of thousands of variables and our code is available~\cite{code}. In the next section, we provide a more concrete formulation of this optimization. 


\subsection{Non-Causal Models}
We give a more explicit formulation of Eq.~\ref{eq:LP1} for the special case of non-causal models.
In this case, each variable sequence, $A_{1:T}$ is a mixture of Markov chains with associated transition probabilities that depend on the unknown value of $R_A$. We denote by $\alpha_+ (\alpha_-)$ the probability that $A$ flips from $0 (1)$ to $1 (0)$ at some time step and $\alpha_0 = P(A_1 = 0)$. We have similar parameters for $B: \beta_{+,-,0}$. We use just $\alpha$ or $\beta$ when possible to avoid writing out all three.
\begin{align}\label{eq:marginal}
\lefteqn{q_A(\alpha) \equiv P(A_{1:T} | R_A) = \alpha_0^{1-A_1} (1-\alpha_0)^{A_1}} \\
&& \alpha_+^{F_{01}(A)} \alpha_-^{F_{10}(A)} (1-\alpha_+)^{F_{00}(A)}  (1-\alpha_-)^{F_{11}(A)}  \nonumber 
\end{align}
The same equations hold replacing $A$ with $B$ and $\alpha$ with $\beta$. $F_{ij}(A)$ counts the number of transitions from state $i$ to $j$ in string $A$. If $A,A'$ have the same initial state and the same transition counts(e.g. $(0,0,1,0)$ and $(0,1,0,0)$), they are said to be {\em partially exchangeable} because they clearly have the same probability of occurring. This observation alone, especially extended to joint strings on $A$ and $B$, imposes serious constraints on possible observed probabilities and explains the existence of equalities like Eq.~\ref{eq:c1}. We discuss tests based on this idea and their relationship to de Finetti theorems in Appendix \ref{sec:appendix}.

In this case, the bounds imposed by Eq.~\ref{eq:bound} are trivial and have already been taken into account by eliminating the dependence of $B_t$ on $A_{t-1}$ in defining the variable above. 
This leaves us with only 12 inequality constraints to enforce positivity, two for each variable: $g_1(\alpha,\beta) = \alpha_0 \geq 0, g_2(\alpha,\beta) = 1- \alpha_0 \geq 0, \ldots, g_{11} (\alpha,\beta) = \beta_+ \geq 0, g_{12} (\alpha,\beta) = 1-\beta_+ \geq 0$. Using $f_{AB} (\alpha,\beta) = q_A(\alpha) q_B(\beta)$ and these definitions for $g_i(\alpha, \beta)$, we can plug these into Eq.~\ref{eq:LP1} to search for bounds that are satisfied by probability distributions explained by non-causal models, $\mathcal P_0$.

\subsection{A Sample Bound}\label{sec:bounds}

As a simple example of how we can use LPs to give bounds, we begin by bounding $P(A=(0,0,1))$ for $P \in \mathcal P_0$. We see that 
$$P(A=(0,0,1)) \leq \max_{\alpha_0,\alpha_+ \in [0,1]} \alpha_0 (1- \alpha_+) \alpha_+$$
 from Eq.~\ref{eq:marginal} and Eq.~\ref{eq:convexity}.  We are looking for an upper bound $\gamma$ so that 
$$
\gamma - \alpha_0 (1- \alpha_+) \alpha_+ \geq 0 \text{ on }\mathcal K,
$$
with $\mathcal K = \{\alpha_0,\alpha_+: \alpha_0,\alpha_+, (1-\alpha_+), (1-\alpha_0) \geq 0 \}$.  Casting the problem as in Eq.~\ref{eq:LP1} with $d_{max}=3$ (with $\vc c$ a constant in this case), we get an LP with $36$ variables (counting $\gamma$) whose solution results in the following representation.
\begin{align*}
\lefteqn{\frac13 - \alpha_0 (1- \alpha_+) \alpha_+ =}\\
&& \frac13(1-\alpha_+)^3+\frac13 \alpha_+^3 + (1-\alpha_0)(1-\alpha_+) \alpha_+
\end{align*}
The RHS constitutes an algebraic proof that $\gamma=1/3$ is an upper bound for $P(A=(0,0,1))$ on this domain. 
Although in principle we can generate algebraic proofs for all bounds presented in the paper, they are unwieldy for all but the simplest examples. Instead we provide code to generate bounds \cite{code}. 

\subsection{Equality Constraints}\label{sec:equality}

A special case of Eq.~\ref{eq:LP1} occurs when we set $d_{max} = 0$. Essentially, we are looking for $\vc c$, so that $\vc c \cdot \vc f (\vc x) = 0$, and we do this by ensuring that the coefficient of each monomial order $x_1^{k_1} x_2^{k_2}\ldots$ equals zero. Putting the constraints from all these coefficients together leads to some matrix $M$ so that $M \vc c = \vc 0$. If T=3, the null space of $M$ has dimension 0, but if T=4 and we consider $\mathcal P_0$, (so that $\vc f(\vc x)$ is set by Eq.~\ref{eq:marginal}), the null space of $M$ has a dimension of 60. This implies that there are 60 linearly independent equalities like the one in Eq.~\ref{eq:c1} that are satisfied by each probability distribution $P(A,B|E) \in \mathcal P_0$.  
A 28-dimensional subspace of these equalities are satisfied for any $\mathcal P_\Delta$. For $T=4$, all of these equalities can be understood in terms of partial exchangeability, see Appendix~\ref{sec:appendix}. Alternately, we can look for particular useful equalities like Eq.~\ref{eq:c1}, or another considered in Appendix~\ref{sec:appendix2}.

In the next section, we focus on inequalities satisfied by all distributions in $\mathcal P_0$, so that violation of these inequalities can rule out a non-causal model. 

\section{Results}\label{sec:results}

Shalizi and Thomas (ST) illustrate 
how the confounding effects of latent homophily cause standard tests for contagion to fail
with two examples~\cite{cosma}. In the first, they demonstrate a non-causal model that looks like contagion.
In the second, they consider a simple copying model and show how the observed results appear to be explained by homophily. Our tests correctly identify the underlying mechanism in both cases.

\subsection{Homophily Looks Like Contagion}
A now popular test for contagion considers unreciprocated, directed edges so that $A$ can influence $B$, but not vice versa~\cite{obesity}. Then, if we regress $B$'s action based on $A$'s history, versus regressing $A$'s action based on $B$'s history, we should see an asymmetry in the size of the regression coefficient if $A$ influences $B$. ST's example shows that this asymmetry can be reproduced by latent homophily as long as there is an asymmetry in the edge formation mechanism. E.g., consider all nodes to take some static hidden attribute in the range $[0,1]$. We say that nodes are more likely to form links with someone who has a similar attribute (homophily), but they also tend to prefer people whose attribute is closer to the median, $0.5$, leading to an asymmetry in preference of edge formation. If each node's state at each time step only depends on their hidden attribute and their previous state, we have a model which clearly has no influence. However, ST show that this model does reproduce asymmetries in regression coefficients which would be interpreted as a sign of influence. 

We ran the code that ST provided in their paper \cite{cosma}, making only one change so that the state of each node at each time step is a binary variable. For a given graph, we consider all pairs of nodes, $A,B$ so that there is a directed edge from $A$ to $B$. Then we look at the frequency of observing a given joint sequence of states for $A_{1:4},B_{1:4}$, and we use this to construct the empirical probability distribution, $\hat P_{LH} (A,B|E)$. We estimated this distribution based on $M=400,000$ samples. As a first test, we can consider the equality constraints that should be satisfied for any non-causal model, given by Eq.~\ref{eq:c1}, Sec.~\ref{sec:appendix2}. 
\begin{align*}
\langle c^{(1)} \rangle_{\hat P_{LH}}&=& (+1499-1493)/400000 \\
 \langle c^{(2)} \rangle_{\hat P_{LH}} &=& (+20006-19871)/400000  
\end{align*}
For a non-causal model we expect exactly 0, but for an empirical distribution the results are not exact due to error in our sampled distribution. In this case, we can calculate a simple confidence bound. Because $c^{(1)}, c^{(2)}$ take only the values $0,\pm 1$, and we are trying to determine if the mean value is nonzero, we can use the binomial distribution to give the exact probability of getting an excess of $+1$(heads) over $-1$(tails) for a fair coin. The p-values we get for the statistics above are 0.54 and 0.25, respectively, which is not extreme enough to rule out the null hypothesis that $\hat P_{LH} \in \mathcal P_0$. 

\begin{figure}[htbp] 
   \centering
   \includegraphics[width=0.8\columnwidth]{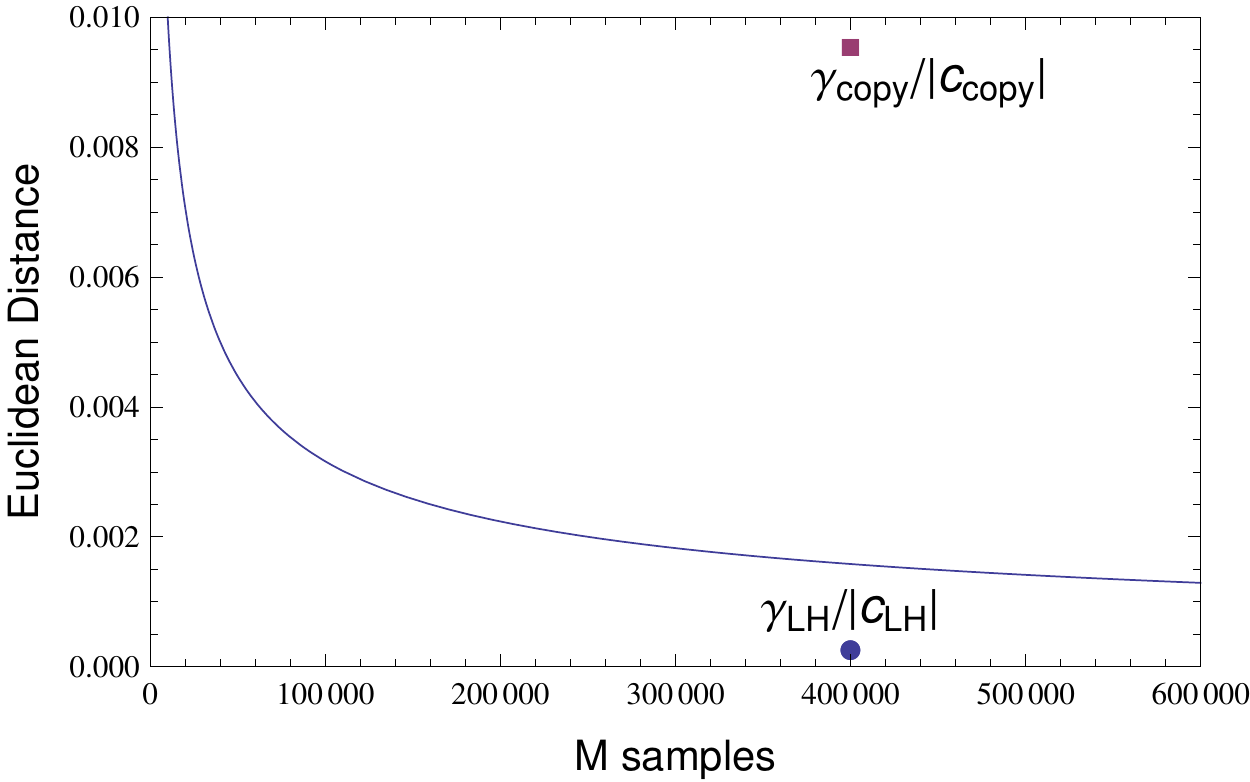} 
   \caption{Average Euclidean distance between the vector of empirical frequencies estimated with $M$ samples and the vector of true probabilities from which the samples were drawn. Compared is the Euclidean deviation from $\mathcal P_0$ by two empirical distributions.}
   \label{fig:deviation}
\end{figure}

The previous test is only one of many conditions we expect non-causal models to satisfy. A more comprehensive test is to take the empirical distribution, $\hat P_{LH}$ and plug it in to Eq.~\ref{eq:LP1} (we used $d_{max}=9$). The result is an observable $c_{LH}(A,B)$, $\gamma_{LH} = 0.0024$, so that $\forall P\in \mathcal P_0, \langle c_{LH} \rangle_{\hat P_{LH}} - \langle c_{LH} \rangle_{P} \geq 0.0024$. How can we interpret this result? Because we have optimized our test based on the data, we cannot apply a straightforward confidence bound like Hoeffding's inequality. 
Instead, we want to check that the Euclidean distance between the empirical distribution and any distribution in $\mathcal P_0$ (which is at least $\gamma/|\vc c_{LH}|$) is much larger than we would expect from sampling error. 
Based on the central limit theorem (and verified by numerical experiments), we expect the average Euclidean distance between a probability distribution and an empirical distribution estimated based on $M$ samples to be $1/\sqrt M$. 
In Fig.~\ref{fig:deviation}, we see that the lower bound on the Euclidean distance between $\mathcal P_0$ and $\hat P_{LH}$, which is $\gamma/|\vc c_{LH}| = 0.00022$ is much less than we would expect from error in sampling the empirical distribution, $\approx 0.0016$, so once again we fail to reject the null hypothesis that $\hat P_{LH} \in \mathcal P_0$.

\subsection{Contagion Looks Like Homophily}

\begin{figure}[tbp] 
   \centering
   \includegraphics[width=0.65\columnwidth]{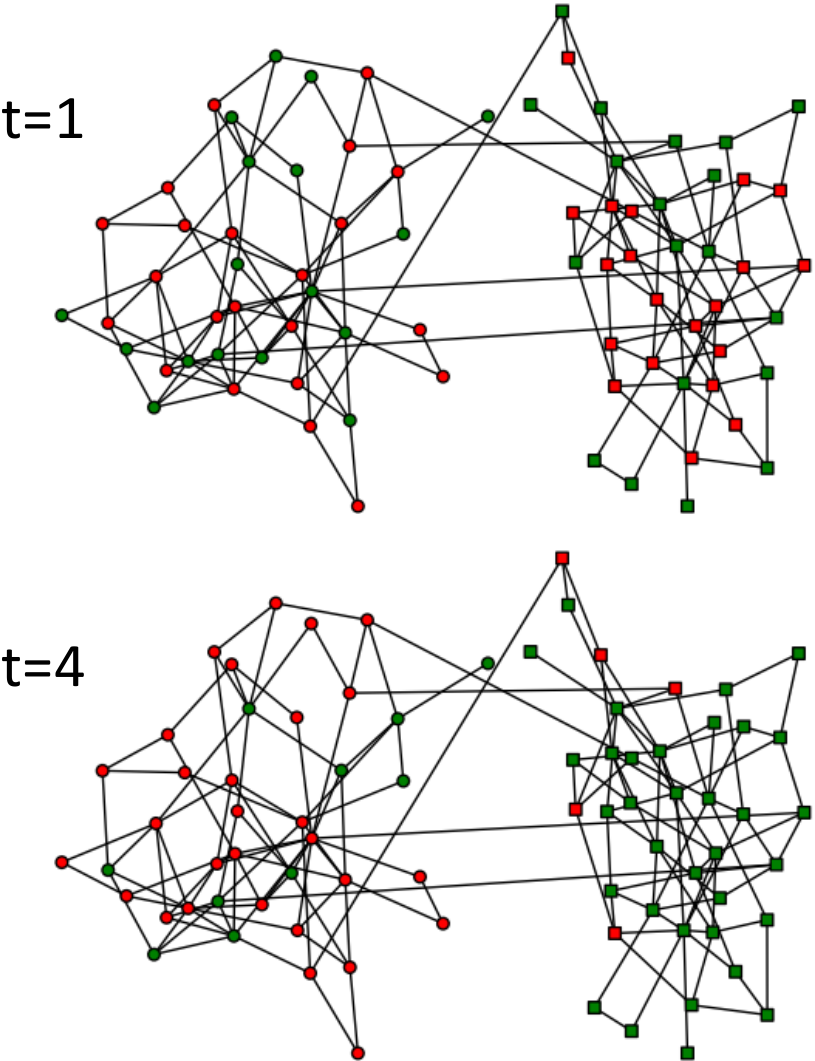} 
   \caption{An example of simple copying dynamics on a network in which color (red/green) spuriously appears to evolve according to the static node type (circle/square). }
   \label{fig:copy}
\end{figure}

In this case, we construct a simple network (see Fig.~\ref{fig:copy}), where each node is defined by a static trait, square or circle. Links are more likely between nodes of the same type. We start each node randomly in the state red or green. We then evolve the state of the graph by repeatedly picking an edge and then copying the state of one node to its neighbor. After many iterations, we observe the new state of the graph. ST point out that by looking at the dynamics it can appear that a tendency to become green, e.g., is explained by the static attribute of being a square, while circles tend to become red. However, this is just transient behavior caused by the network structure, the node type has nothing to do with the copying mechanism. 

Again, we generate $M=400,000$ samples using this model (details and code in \cite{cosma}), generating an empirical distribution, $\hat P_{copy}$. Our test easily identifies contagion in this case. For instance, 
$\langle c^{(1)} \rangle_{\hat P_{copy}} =  0.062,$
for which the p-value under the null hypothesis is $\sim 10^{-3000}$. If, on the other hand, we solve Eq.~\ref{eq:LP1} using $\hat P_{copy}$, we get $\gamma_{copy}/|c_{copy}| = 0.0095$, which is much larger than we would expect from error in sampling the empirical distribution ($0.0016$, see Fig.~\ref{fig:deviation}). 


\subsection{Experimental Results}\label{sec:experimental}

We have done a preliminary analysis of the Framingham Heart Study (FHS) data, a longitudinal social network study which includes many covariates (e.g. obesity, marriage, depression, smoking, alcohol) and link types (friend, neighbor, co-worker, sibling, spouse). Detailed analysis of this dataset appears in several works~\cite{obesity,christakis2012}. Methodologically, we proceed in an identical fashion to the previous sections.

We consider a particular example from the Framingham Heart Study regarding BMI (a BMI greater than 30 is defined as ``obese''). We considered waves 4,5,6, and 7 of the original and offspring cohort, and considered pairs $A,B$ where $B$ nominated $A$ as a friend, and $A$ and $B$ are not related. Because our goal is to rule out the non-causal model, the timing of the edge creation is not a factor. We defined the binary variable $A_t = 0 (1)$ to indicate that $A$'s BMI did not (did) increase by more than the median amount since the last survey. This definition was intended to reduce the effect of dynamic factors influencing all actors in the same way. The median change in BMI for these 4 waves was $(0.55, 0.57, 0.42, 0.20)$. As we mention in the conclusion and Appendix \ref{sec:appendix}, we can also test whether the data is consistent with a model including latent homophily and unbounded contagion as a way of identifying the presence of other unmodeled causal effects. We were not able to rule out that $\hat P_{BMI} \in \mathcal P_\Delta$ (which would suggest the presence of unmodeled causal effects), but we could rule out $\hat P_{BMI} \in \mathcal P_0$ with high confidence. 
For instance, using the test in Eq.~\ref{eq:c1}
we can rule out a non-causal model with $99\%$ confidence. 
Since even this claim is controversial, we have included the relevant statistics for observing joint sequences of actions of friends in the supplementary material~\cite{code}, so that the violation of Eq.~\ref{eq:c1} can be verified directly. 

As a contrast, let $A_t = 0 (1)$ indicate that Alice smoked at least one cigarette per day. In that case, we could rule out $\hat P_{smoking} \in \mathcal P_\Delta$ with high confidence for any $\Delta$. Even latent homophily combined with contagion do not explain the correlations in smoking. 
A complete analysis of FHS data will appear in future work. 

%
%
%
%
%
%

\section{Related work}\label{sec:related}

Christakis and Fowler were not the first to look for contagion in observational social network studies, but their study on obesity~\cite{obesity} marks the beginning of an eruption of methodological introspection and critique which CF have recently summarized and addressed~\cite{christakis2012}. While most of the responses to that work center around the robustness of various parametric modeling assumptions (e.g., sensitivity analysis~\cite{vanderweele}), our central concern is with the broader, so far unanswered, critique leveled by ST that latent homophily poses a significant barrier to identifying contagion~\cite{cosma}. 

The main difference to ST's paper is that instead of identifying the strength of causal effects, we put lower bounds on these effects (which would seem to be the primary aim for most practitioners). In this sense, we are actually in the realm of ``partial identifiability''~\cite{manskibook}, an approach ST themselves suggest as an open problem in the final section of their paper~\cite{cosma}. 

The approach we have taken here considers probability distributions as points in some high-dimensional vector space. In this context we can apply tools from algebraic geometry to answer a variety of questions about set membership. While the possibility of this approach was recognized long ago \cite{geigermeek,kangtian}, those approaches relied on computationally infeasible, exact methods. Recent advances in convex relaxations for algebraic geometry problems~\cite{parrilo} make this approach feasible. 
This perspective was considered in a general context~\cite{versteeg2011uai}, but the result here differs in important ways. Casting the optimization problem here as an LP instead of a semi-definite program eliminates some ambiguities in defining an optimal test, leads to a more tractable optimization, and allows us to address more complex problems. Ultimately, this added power allows us to solve a concrete, open problem in causal analysis by bounding the strength of contagion even in the presence of confounders. 
\section{Discussion}

When it comes to human behavior, measuring and controlling for every variable that might be relevant is unrealistic. 
For that reason, it is important to obtain the best possible bounds on causal strength with the fewest possible assumptions.
In the context of a specific graphical model like Fig.~\ref{fig:homophily}, we can unambiguously state our result as a lower bound on the strength of contagion. 

In the real world, contagion is only one of many effects that could cause deviation from the non-causal model. For example, common external causation could be a factor. 
Even in that situation our test can provide valuable insight.
For instance, we can test whether a distribution belongs to the set of models which include latent homophily and unbounded contagion strength. If not, we can conclude that other factors must be involved. In our analysis of FHS data, we found this to be the case for smoking, for example. This is not surprising if changing laws and restrictions on public smoking act as a common external cause of correlations. 

The methods outlined in this paper provide a powerful way to statistically test for causal effects, even in the presence of confounding variables, while invoking minimal assumptions. 
While this solves the open problem of distinguishing latent homophily and contagion, our analysis highlights another widespread shortcoming of social network studies. 
Assuming that no external dynamics influence correlations, while a common assumption, is often unjustified. The methods we present here can also be used to test the validity of this assumption. 
Future work will apply these tests to an in-depth analysis of causal effects in the Framingham Heart Study data.

\appendix
\section{Joint Partial Exchangeability}\label{sec:appendix}

Consider the set of pairs  $z,z' \in \{0,1\}^T$ that obey the relation $z \equiv_e z'$. Here $\equiv_e$ is an equivalence relation that denotes partial exchangeability (PE). Two sequences are partially exchangeable if and only if they have the same initial state and the same count for each transition (which we called $F_{ij} (Z)$ in the text). Clearly, a stationary Markov chains produces partially exchangeable sequences with equal probability, because the probabilities depend only on the initial state and number of transition. Even if we average over Markov chains with unknown, arbitrary transition matrices, this property is preserved. 
A famous result known as the de Finetti theorem for Markov chains is that in the limit of long sequences, PE (along with a small technical requirement that there is always a probability of visiting some state) is also a sufficient condition for data to be described as a mixture of Markov chains~\cite{diaconis}. Essentially, we have a theorem proving that a symmetry property justifies a particular form for a latent variable graphical model.

\begin{figure}[htbp] 
   \centering
   \includegraphics[width=0.9\columnwidth]{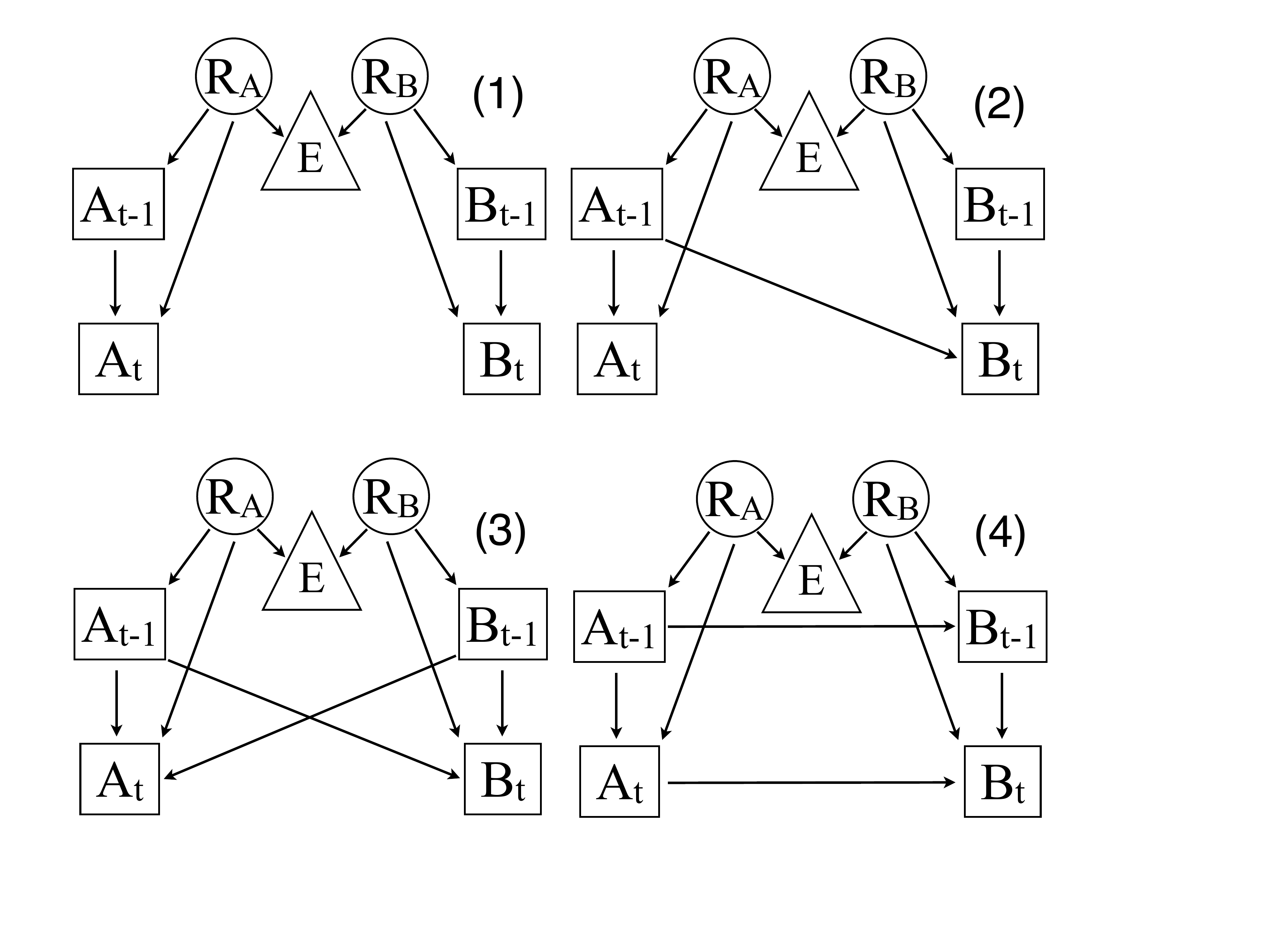} 
   \caption{Four model variations with different exchangeability properties.}
   \label{fig:allmodels}
\end{figure}

In the context of the null model, $\mathcal P_0$ (model 1 in Fig.~\ref{fig:allmodels}), where we ask whether Alice and Bob's sequences are both described by mixtures of Markov chains this suggests a simple test. For each joint sequence $A,B$ we should see that $P(A,B) = P(A',B')$ if $A \equiv_e A'$ and $B \equiv_e B'$. We call this notion joint partial exchangeability (JPE), which differs from PE, because statistics for Alice and Bob's sequences could both be PE while failing to be JPE (i.e. $P(A)=P(A'), P(B)=P(B'), P(A,B)\neq P(A',B')$). As an example of the different possibilities consider the sequences $w=(1,0,0,0), x = (0,0,1,0),  y =(0,1,0,0), z = (0,0,0,1)$. Clearly, $x \equiv_e y$. Consider also the alternative models in Fig.~\ref{fig:allmodels}. We can view model (3), for instance, as a mixture of stationary Markov chains on the joint variable $(A_t,B_t)$, but this implies different exchangeability properties. Going back to our example, we see the difference.
{
\scriptsize
\begin{tabular}{ccc}
Equality & Models Satisfying \\\hline
$P(A = x, B=x) = P(A = y, B=y)$ &(1),(2),(3),(4) \\
$P(A = w, B=x) = P(A = w, B=y)$ &(1),(4) \\
$P(A = x, B=x) = P(A = x, B=y)$ &(1) \\
$P(A = z, B=x) = P(A = z, B=y)$ &(1),(2) 
\end{tabular}
}

For longer sequences of observations, this type of test would be computationally much easier than solving the optimization suggested in Eq.~\ref{eq:LP1}, which is exponential in $T$. De Finetti's theorem suggests that in the long $T$ limit, this test may even be sufficient. 

This test has another nice interpretation in terms of a symmetry condition. If there is influence from $A$ to $B$ some (but not all) of the symmetries are eliminated (see Sec.~\ref{sec:equality}). If there is influence in both directions, an even smaller (but nonzero) set of symmetries remain. Essentially, model checking then becomes a question of (approximately) matching symmetries in the data to the appropriate models. Interestingly, these symmetries differ from typical conditional independence relations. Because we have a mixture of Markov processes, none of the observed variables are independent. A similar perspective has been applied for testing the order of a Markov process~\cite{quintana}.

\section{Another Equality}\label{sec:appendix2}
\renewcommand{\tabcolsep}{1mm}
\begin{table}[htbp]
\centering
{
\tiny
\begin{tabular}{cccccccccccccccc}
0& 0& -1& 0& 1& 0& 0& 0& 0& 0& 0& 1& 0& -1& 0& 0\\
0& 0& 1& 0& -1& 0& 0& 0& 0& 0& 0& 1& 0& -1& 0& 0\\
1& 1& -1& -1& 1& 1& 1& 1& -1& -1& -1& -1& 1& 1& 1& -1\\
0& 0& -1& 0& 1& 0& 0& 0& 0& 0& 0& 1& 0& -1& 0& 0\\
-1& -1& 1& 1& -1& -1& -1& -1& 1& 1& 1& 1& -1& -1& -1& 1\\
0& 0& 1& 0& -1& 0& 0& 0& 0& 0& 0& 1& 0& -1& 0& 0\\
0& 0& -1& 0& 1& 0& 0& 0& 0& 0& 0& -1& 0& 1& 0& 0\\
0& 0& -1& 0& 1& 0& 0& 0& 0& 0& 0& 1& 0& -1& 0& 0\\
0& 0& 1& 0& -1& 0& 0& 0& 0& 0& 0& -1& 0& 1& 0& 0\\
0& 0& 1& 0& -1& 0& 0& 0& 0& 0& 0& 1& 0& -1& 0& 0\\
0& 0& -1& 0& 1& 0& 0& 0& 0& 0& 0& -1& 0& 1& 0& 0\\
-1& 1& 1& 1& -1& 1& 1& 1& -1& -1& -1& 1& 1& -1& -1& 1\\
0& 0& -1& 0& 1& 0& 0& 0& 0& 0& 0& 1& 0& -1& 0& 0\\
1& -1& 1& -1& -1& -1& -1& -1& 1& 1& 1& 1& -1& -1& 1& -1\\
0& 0& -1& 0& 1& 0& 0& 0& 0& 0& 0& -1& 0& 1& 0& 0\\
0& 0& -1& 0& 1& 0& 0& 0& 0& 0& 0& 1& 0& -1& 0& 0
\end{tabular}} 
\caption{Specifying an observable for which $\forall P \in \mathcal P_0, \langle c^{(2)}(A,B) \rangle_P = 0$.}\label{eq:c2}
\end{table}
We can look for equalities that are useful for identifying certain types of influence by solving Eq.~\ref{eq:LP1} using some influence model for $\hat P$ and setting $d_{max}=0$. We already saw an example  in Eq.~\ref{eq:c1} which was obtained by setting $T=4$ and defining $P_\delta^{(i)} (A,B)$ so that $P(A_t=0) =P(A_t=1)= 1/2$ and $B_t = A_{t}$ with probability $\delta$ otherwise $B_t$ randomly becomes $0$ or $1$. This can be thought of as an ``instant'' influence model, where Alice's choice at time $t$ influences Bob's choice at time $t$. In contrast, we had previously defined a delayed influence model, $P_\delta (A,B)$ where $B_t = A_{t-1}$ with probability $\delta$. 
Solving Eq.~\ref{eq:LP1} using $P_\delta$ leads to another equality constraint, $\forall P \in \mathcal P_0$, $\langle c^{(2)} \rangle_{P} =0$, listed in Table~\ref{eq:c2}.
For models with influence, on the other hand, we have the following. 
\begin{align*}
\langle c^{(2)} \rangle_{P_\delta}&=& \frac7{16} \delta  \qquad
 \langle c^{(1)} \rangle_{P_\delta^{(i)}} &=&\frac18 (3-\delta^2) 
\end{align*}
The value of $c^{(2)}(A,B)$, for a particular $A \in \{0,1\}^4, B \in \{0,1\}^4$ can be read from Table~\ref{eq:c2}. Simply consider the sequence $A$ or $B$ as a binary number and pick the $A$-th row and the $B$-th column.  This equality is the maximally violated one for a particular model of delayed influence, but it provided significant violations for synthetic and real-world data in Sec.~\ref{sec:results}.

\subsubsection*{Acknowledgements}
We are grateful to Nicholas Christakis for providing access and advice for the FHS data. 
GV thanks Michael Finegold and Andrew C. Thomas for useful conversations while visiting LARC, SMU.
This work was supported under AFOSR grant FA9550-12-1-0417 and AFOSR MURI grant No. FA9550-10-1-0569.

%

\end{document}